\documentclass[pra,aps,reprint,superscriptaddress,amsmath,amssymb]{revtex4-1}
\usepackage{graphicx,bm,hyperref,color,ulem}
\usepackage[caption=false]{subfig}
\usepackage{mathtools}
\usepackage[utf8]{inputenc}
\usepackage[T1]{fontenc}
\begin{document}
\title{Polarization tagging of two-photon double ionization by elliptically polarized XUVs }
\author{Stefan Donsa}
\email{stefan.donsa@tuwien.ac.at}
\affiliation{Insitute for Theoretical Physics, Vienna University of Technology, A-1040 Vienna, Austria, EU}
\author{Iva B\v{r}ezinov\'a}
\affiliation{Insitute for Theoretical Physics, Vienna University of Technology, A-1040 Vienna, Austria, EU}
\author{Hongcheng Ni}
\email{hongcheng.ni@tuwien.ac.at}
\affiliation{Insitute for Theoretical Physics, Vienna University of Technology, A-1040 Vienna, Austria, EU}
\author{Johannes Feist}
\affiliation{Departamento de F\'isica Te\'orica de la Materia Condensada and Condensed Matter Physics Center (IFIMAC), Universidad Aut\'onoma de Madrid, E-28049 Madrid, Spain, EU}
\author{Joachim Burgd\"orfer}
\affiliation{Insitute for Theoretical Physics, Vienna University of Technology, A-1040 Vienna, Austria, EU}
\begin{abstract}
We explore the influence of elliptical polarization on the (non)sequential two-photon double ionization of atomic helium with ultrashort extreme ultraviolet (XUV) light fields using time-dependent full \textit{ab-initio} simulations.
The energy and angular distributions of photoelectrons are found to be strongly dependent on the ellipticity.
The correlation minimum in the joint angular distribution becomes more prominently visible with increasing ellipticity.
In a pump-probe sequence of two subsequent XUV pulses with varying ellipticities, polarization tagging allows to discriminate between sequential and nonsequential photoionization.
This clear separation demonstrates the potential of elliptically polarized XUV fields for improved control of  electronic emission processes.
\end{abstract}

\maketitle

\section{Introduction}

With the advent of high-order harmonic generation (HHG) \cite{KraSchKul1992,MacKmeGor1993,PopCheAr2010} extreme ultraviolet (XUV) pulses are now routinely generated in table-top experiments making the studies of distinct aspects of light-matter interaction possible. 
Recent progress in HHG from bichromatic counter-rotating circularly polarized laser pulses \cite{FleKfuDis2014,Mil2015,BakAhsLin2016,MauBanUze2016,ReiMad2016,MedWraHar2015,KfiGryTur2015,FanGryKnu2015,CheTaoHer2016,HuaHerHua2018} and from relativistic laser driven plasma mirrors \cite{ChePuk2016,MaYuYu2016} has opened the door to generation and application of elliptic XUV pulses giving an additional knob to study and, most importantly, to control strong-field atomic, molecular, and surface dynamics. 
Elliptically polarized energetic pulses can also be produced with free-electron lasers \cite{AllDivCal2014,LutMaxIlc2016,KorWedNol2017}, extending the study to inner-shell electronic dynamics. 
The level of control which can be achieved was recently demonstrated by producing a spiral pattern in the momentum distribution of ionized electrons which were produced by two oppositely circularly polarized time-delayed XUV pulses \cite{NgoHuMad2015,PenKerJoh2017}. \\
%These studies have also been extended to two-photon double ionization in helium \cite{NgoManMer2014,NgoSta2017}.\\
%In this work we show how elliptically polarized light fields can be used to understand more details of the interplay between sequential and non-sequentially processes in two-photon double ionization.
Two-photon double ionization (TPDI) of atomic helium is a prototypical process to investigate electron correlations and has been studied extensively using linear XUV pulses, e.g. \cite{FeiNagPaz2008,FeiNagPaz2009,PazFeiNag2011,FeiNagTic2011,PazNagBur2015,LauBac2003,ZhaPenXu2011,HuColCol2005,KheIva2006,ZhaPenXu2011,PalResMcc2009,PalHorRes2010,PalResMcc2009b}, and recently also with elliptically polarized light fields \cite{NgoManMer2014,PinLiCol2016,NgoSta2017}.
For photon energies below the second ionization potential $I_2=54.4$ eV, the system is in the spectrally nonsequential regime where TPDI only occurs for near-simultaneous absorption of two photons even for long pulses.
For higher photon energies, the two-photon double continuum can be reached by two sequential one-photon absorptions via on-shell intermediate states of the He$^{+}$ ion.
For ultrashort pulses the clear distinction between nonsequential and sequential regimes breaks down.
Even for XUV pulses in the spectrally sequential regime with central energies above $I_2$, TPDI becomes nonsequential in the time domain as the first electron is still in the vicinity of the ion when the second absorption process takes place.
Exchange of energy and angular momentum between the departing electrons, referred to in the following as dynamical Coulomb correlation, persists which strongly depends on the duration of the ionizing pulse \cite{FeiNagPaz2009,PalHorRes2010}.\\
In the present article we demonstrate that in addition to the pulse duration also the ellipticity $\epsilon$ of the ultrashort XUV pulse provides an effective knob to control the TPDI process. 
We show that the dynamical correlation features in the joint angular distribution become more clearly identifiable and more pronounced for circularly polarized fields compared to linearly polarized light fields as distortions due to dipolar nodal planes are absent. 
Moreover, for a sequence of two ultrashort pulses with ellipticities $\epsilon_1$ and $\epsilon_2$, multi-path interferences in the double continuum can be controlled.
Polarization tagging allows to distinguish between electron emission by the first or second of the two pulses.
We explore possible experimental signatures accessible with a reaction microscope \cite{UllMosDor2003} as recently demonstrated for ionization in the strong-field multi-photon double ionization regime \cite{HenEckHar2018_1,HenEckHar2018_2}.

\section{Numerical Method}
We solve the time-dependent Schr\"odinger equation (TDSE) for atomic helium in its full dimensionality including the full electron-electron interaction by expanding the wave function in coupled spherical harmonics
\begin{equation}
\Psi \left(\bm{r}_1,\bm{r}_2,t \right)=\sum_{L,M}^{\infty} \sum_{l_1,l_2}^{\infty} \frac{R_{l_1,l_2}^{L,M}(r_1,r_2,t)}{r_1 r_2 }\mathcal{Y}_{l_1,l_2}^{L,M}\left( \Omega_1, \Omega_2 \right),
\end{equation}
with
\begin{eqnarray}
& \mathcal{Y}_{l_1,l_2}^{L,M}\left( \Omega_1, \Omega_2 \right) = \nonumber \\
& \sum\limits_{m_1,m_2}\left\langle l_1 m_1 l_2 m_2 | l_1 l_2 L M \right\rangle Y_{l_1,m_1}\left( \Omega_1 \right)Y_{l_2,m_2}\left( \Omega_2 \right).
\end{eqnarray}
Inserting this expansion into the TDSE yields the time-dependent close coupling (TDCC) equations \cite{PinRobLoc2007,LauBac2003,ColPin2002,HuColCol2005}.
The electric fields are treated in dipole approximation.
In the present work we have extended our TDCC approach \cite{FeiNagPaz2008} previously limited to the case of linearly polarized light fields to arbitrarily polarized fields.
Accordingly, the total magnetic quantum number $M$ is no longer a conserved quantity.
This substantially increases the number of close-coupling equations for the same maximum total angular momentum $L_{\mathrm{max}}$ from $(L_{\mathrm{max}}+1)$ to $(L_{\mathrm{max}}+1)^2$ and thereby the numerical complexity.
The radial wave functions $R_{l_1,l_2}^{L,M}(r_1,r_2,t)$ are discretized on a spatial grid using the finite-element discrete-variable-representation (FEDVR) \cite{ResMcc2000,MccHorRes2001,SchCol2005,SchColHu2006}.
For the temporal propagation we use the short-iterative Lanczos algorithm with adaptive time step control \cite{ParLig1986,SmyEdwPar1998,LefBisCer1991}.
To extract the spectral information we project the six-dimensional wave function onto products of uncorrelated energy-normalized Coulomb wave functions for each combination of $L$, $M$, $l_1$, and $l_2$ separately.
For the projection to be converged the wave function has to be propagated sufficiently long after the end the pulses, for a detailed discussion see \cite{FeiNagPaz2008}.
We use velocity gauge throughout and find converged results with a close-coupling expansion of $L_{\mathrm{max}}=$ 3, $l_{1,\mathrm{max}}=l_{2,\mathrm{max}}=$ 12.
\section{Influence of ellipticity and pulse duration  on the angular distribution}
\begin{figure}
\includegraphics[width=1\columnwidth]{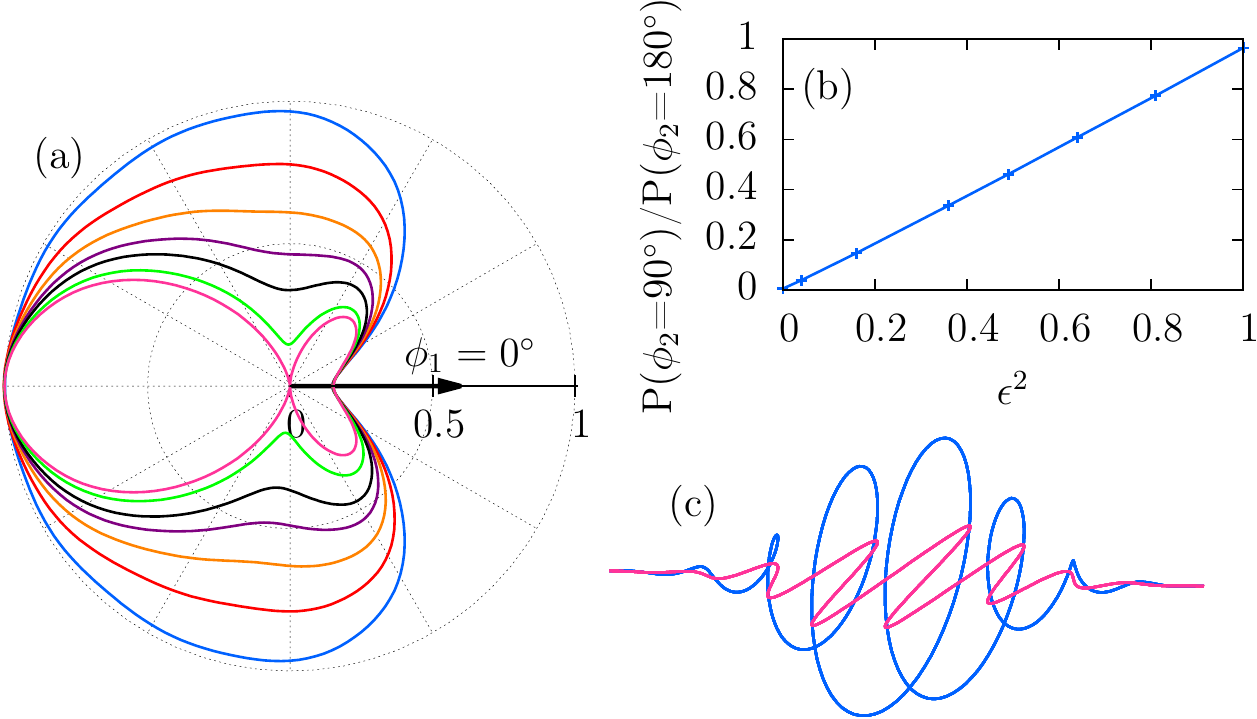} 
\caption{(a) Joint angular distribution $P(\theta_1=\theta_2=$90$^{\circ},\phi_1=$0$^{\circ}, \phi_2)$ of the ejected electrons for a pulse duration $T_p=$ 150 as at $\hbar \omega = $70 eV for  different ellipticities $\epsilon$. 
The innermost line (pink) corresponds to the previously investigated case of linear polarization along $\hat{x}$ ($\epsilon$= 0), whereas the successive outer curves correspond to $\epsilon=$ 0.4, 0.6, 0.7, 0.8, 0.9, and 1 (circular polarization).
All distributions are normalized to a maximum value of one.
(b) Ratio of probability for emission of the two electrons with a relative angle of 90$^{\circ}$ to the probability for emission into the opposite direction (back-to-back) for the different ellipticities shown in (a).
(c) Sketch of a laser field with $\epsilon$= 1 (blue) and 0 (pink).}\label{fig:tdcs_single_pulse_ellip}  
\end{figure}
\begin{figure}
\includegraphics[width=1\columnwidth]{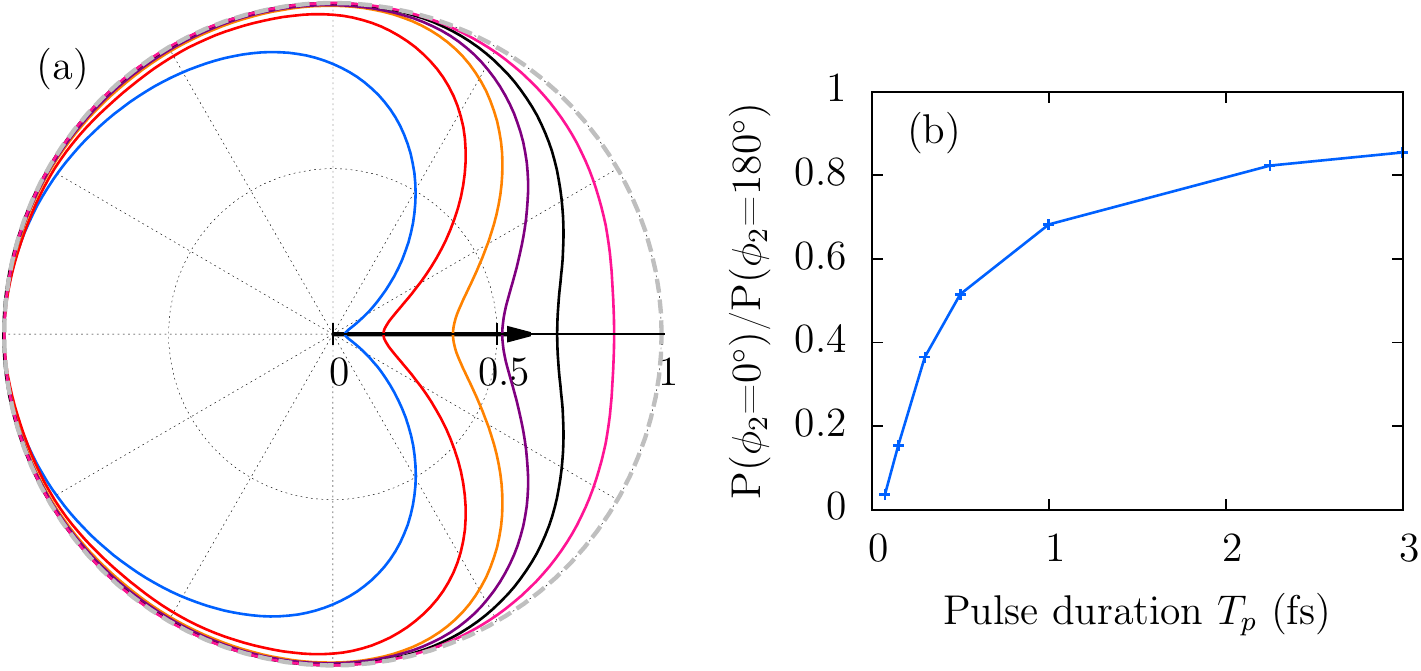} 
\caption{(a) Joint angular distributions $P(\theta_1=\theta_2=$90$^{\circ},\phi_1=$0$^{\circ}, \phi_2)$ of the ejected electrons for different pulse durations at $\hbar \omega = $70 eV.
The innermost line (blue) is for $T_p=$ 75 as, with successive outer lines for $T_p=$ 150, 300, 500, 1000, 3000 as. 
The dashed line is the unit circle. 
All distributions are normalized to the maximum value of the angular distribution at $\phi_2=$ 180$^{\circ}$.
(b) Ratio of emission of the two electrons into the same (side-by-side) and into the opposite direction (back-to-back) for the different pulse durations shown in (a).
 }\label{fig:tdcs_single_pulse_length}  
\end{figure}

We use XUV pulses with central photon energies of 65 and 70 eV, a  peak intensity of $10^{12}$ W/cm$^2$, which is in the purely perturbative regime for these photon energies and a cos$^2$ envelope for the vector potential given by $f(t)=\cos ^2 \left( \frac{\pi}{2 T_p} (t-t_{XUV})  \right)$ for $-T_p<(t-t_{XUV})<  T_p$ where $t_{XUV}$ is the peak time of the XUV and $T_p$ is the full-width at half maximum duration of the vector potential. 
We simulate the response to ultrashort pulses with $T_p\lesssim$ 2 fs corresponding to a Fourier bandwidth $\Delta \omega \approx 2 \pi / T_p >$ 2 eV.
Despite this considerable width, the spectral overlap with the energetically nonsequential regime $\hbar \omega < I_2=$ 54.4 eV is still negligible.
We therefore focus on the nonsequential regime in the time domain.
The elliptical vector potential, propagating along the $\hat{z}$ axis and polarized in the $\hat{x}-\hat{y}$ plane, is defined by
\begin{equation} \label{eq:field}
\vec{A}(t)= A_0 f(t) \begin{pmatrix} \sin\left( \omega (t-t_{XUV})  \right) \\ - \epsilon\cos\left( \omega (t-t_{XUV})\right)\\0\end{pmatrix},
\end{equation}
where $\epsilon$ is the ellipticity of the laser field.
The light field is called left(right)-circularly polarized if $\epsilon$= 1($-$1).
For TPDI by ultrashort XUV pulses electron-electron interactions in the double continuum leave a strong imprint on the energy and angular distribution of the emitted electron pair \cite{FeiNagPaz2009,PazFeiNag2011,PalHorRes2010}.
Angular correlations between the two electrons are characterized by the joint angular distribution
\begin{equation}\label{eq:joint_angular}
P(\theta_1, \theta_2,\phi_1,  \phi_2) = \int \mathrm{d}E_1 \mathrm{d}E_2 P \left(E_1,E_2,\Omega_1,\Omega_2 \right)
\end{equation}
integrated over the energies $E_{1,2}$ of the emitted electrons.
Within the polarization plane $(\theta_1=\theta_2=$ 90$^{\circ})$ and for one electron emitted along the $\hat{x}$-axis, $P($90$^{\circ},$90$^{\circ},\phi_1=$ 0$^{\circ}, \phi_2)$ strongly varies with the ellipticity.
For linear polarization ($\epsilon$=0), $P(\phi_2)$ displays the previously observed "fish-like" angular distribution [Fig.~\ref{fig:tdcs_single_pulse_ellip}~(a)].
With increasing $\epsilon$, the dip due to the nodal line at $\phi$=90$^{\circ}$ disappears.
The ratio $P(\phi_2$=90$^{\circ}$)/$P(\phi_2$=180$^{\circ}$) approximately scales with $\epsilon^2$ [Fig.~\ref{fig:tdcs_single_pulse_ellip}~(b)].
In the limit of circular polarization the only structure remaining is the suppression of electron emission into the same direction ($\phi_2$=0$^{\circ}$) (side-by-side) relative to the back-to-back emission ($\phi_2$=180$^{\circ}$). 
The pronounced minimum for emission of both electrons in the same direction is not affected by the variation of $\epsilon$.
The obvious reason for the strong suppression is the repulsive electron-electron interaction, when the two-electron emission is temporally confined to a fraction of a femtosecond and is a prototypical case of dynamical Coulomb correlation.\\
Performing a scan of the pulse duration $T_p$ \cite{FeiNagPaz2009} we find that with increasing pulse duration the dip in the angular distributions becomes less and less pronounced and in the limit of very long pulses it approaches a purely circular distribution (grey dashed lines), see Fig.~\ref{fig:tdcs_single_pulse_length}~(a).
This dynamical Coulomb correlation can be quantified by the dependence of the ratio $P(\phi_2=$0$^{\circ})/P(\phi_2=$180$^{\circ})$ on $T_p$ [Fig.~\ref{fig:tdcs_single_pulse_length}~(b)].
In the limit $T_p \rightarrow 0$, Coulomb repulsion tends to completely block the side-by-side emission.
For circular polarization, the side-by-side suppression is the dominant structure in the joint angular distribution, while for linear polarization additional pronounced minima due to the nodal plane may overshadow this effect in the experiment.
Circularly polarized XUV's are therefore the preferred experimental setting to unambiguously establish the dynamical correlation in TPDI for ultrashort pulses.
Moreover, unlike for linear polarization the joint angular distribution is rotationally invariant   $P(\phi_1=$ 0$^{\circ},\phi_2)= P(\phi_1=\alpha,\phi_2+\alpha)$ for circular pulses. 
Consequently, the correlation dip will persist when integrating the experimental angular distribution over the azimutal angle $\alpha$ while keeping the relative angle $\phi_1-\phi_2$ fixed, thereby improving the experimental signal-to-noise ratio.
\section{XUV-XUV pump-probe sequence with elliptically polarized pulses}
XUV-XUV pump-probe sequences have been theoretically investigated for linearly polarized pulses in the past, e.g. \cite{PalResMcc2009,PalHorRes2010,FeiNagTic2011,FouAntBac2008,MorWatLin2007,JiaWeiTia2014}.
Palacios et al. \cite{PalResMcc2009,PalHorRes2010} investigated an XUV-XUV sequence for pulses with different energies and varying time delays and found an interference pattern in the angle integrated, but energy resolved double ionization probability $P^{DI}(E_1,E_2)$ as a function of the energy difference $\Delta E=E_1-E_2$ between the electrons.
XUV-XUV pulse sequences have also been used to explore double ionization via doubly excited resonances as intermediate states of interfering pathways \cite{FeiNagTic2011}. 
\begin{figure}
\includegraphics[width=1\columnwidth]{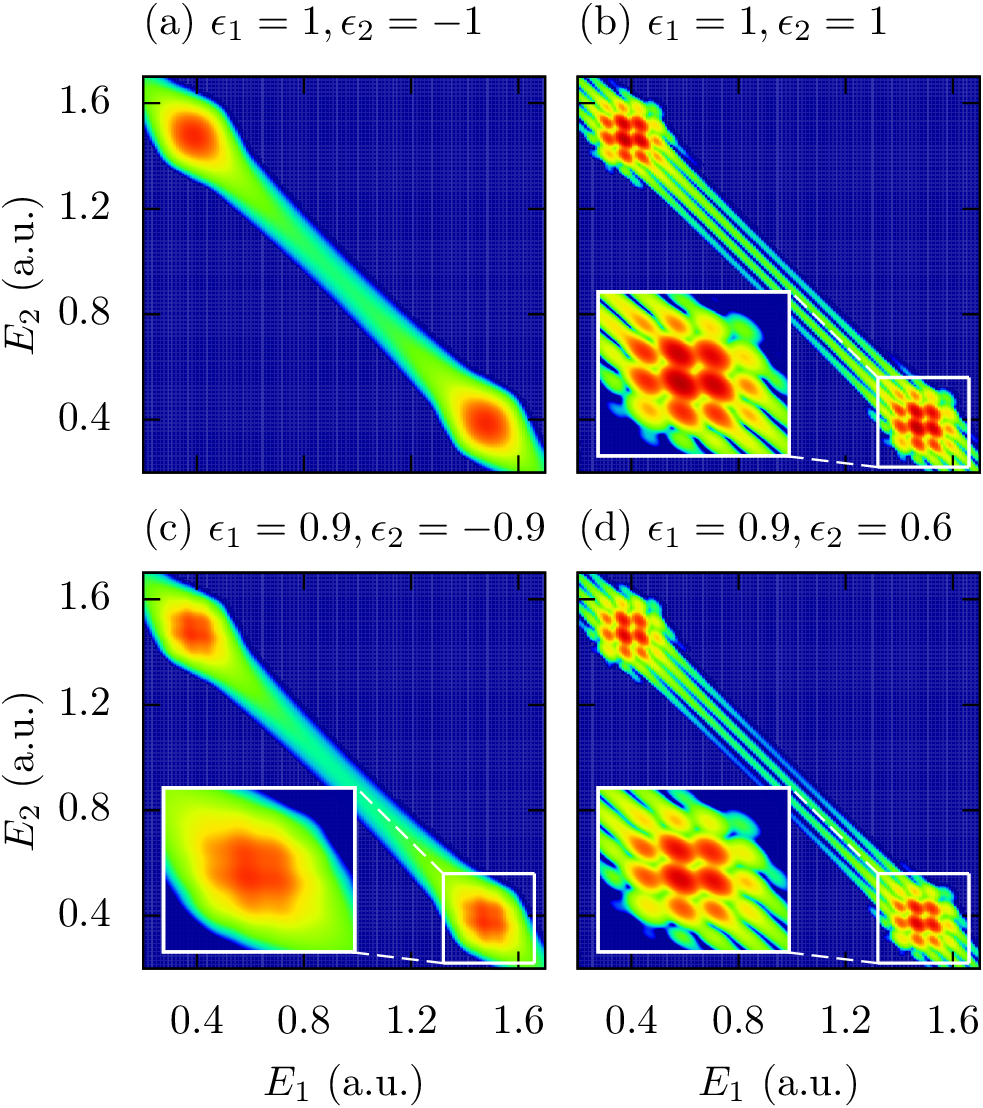} 
\caption{Double ionization spectrum $P^{DI}(E_1,E_2)$ (log-scale) as a function of $E_1$ and $E_2$  for pulse sequences with different ellipticities. 
The central photon energy of both pulses is 65 eV, their duration is $T_p$= 1 fs and the pulse delay $\tau$= 2.5 fs. }\label{fig:pdi_e1_e2}
\end{figure}
\\
We explore in the following double ionization by a sequence of two ultrashort XUV pulses with, in general, different ellipticities $\epsilon_{1,2}$.
We show that polarization tagging allows to distinguish ionization events occurring in the first and the second pulse and, thus, temporal (non)sequential ionization.
With this control knob interferences can be switched on and off.
The two pulses in Fig.~\ref{fig:pdi_e1_e2} have identical central energy of 65 eV and a Fourier width corresponding to $T_p$= 1 fs.
The time delay between the pulses is sufficiently large ($\tau$= 2.5 fs) that the two pulses are temporally well separated.
\subsection{Doubly differential energy distributions}
We first consider the angle-integrated double ionization probability as a function of the two electron energies $P^{DI}\left(E_1,E_2\right)$.
In the $E_1-E_2$ plane we observe two well separated peaks signifying sequential double ionization and a faint ridge connecting the two indicating a weak non-sequential contribution. 
For opposite polarizations [$\epsilon_1$= 1, $\epsilon_2$= $-$1, Fig.~\ref{fig:pdi_e1_e2}~(a)] the peaks are structureless.
By contrast, for two pulses with the same ellipticity  ($\epsilon_1=\epsilon_2=$~1) $P^{DI}(E_1,E_2)$ displays an intricate interference pattern [Fig.~\ref{fig:pdi_e1_e2}~(b)].
\begin{figure}
\includegraphics[width=1\columnwidth]{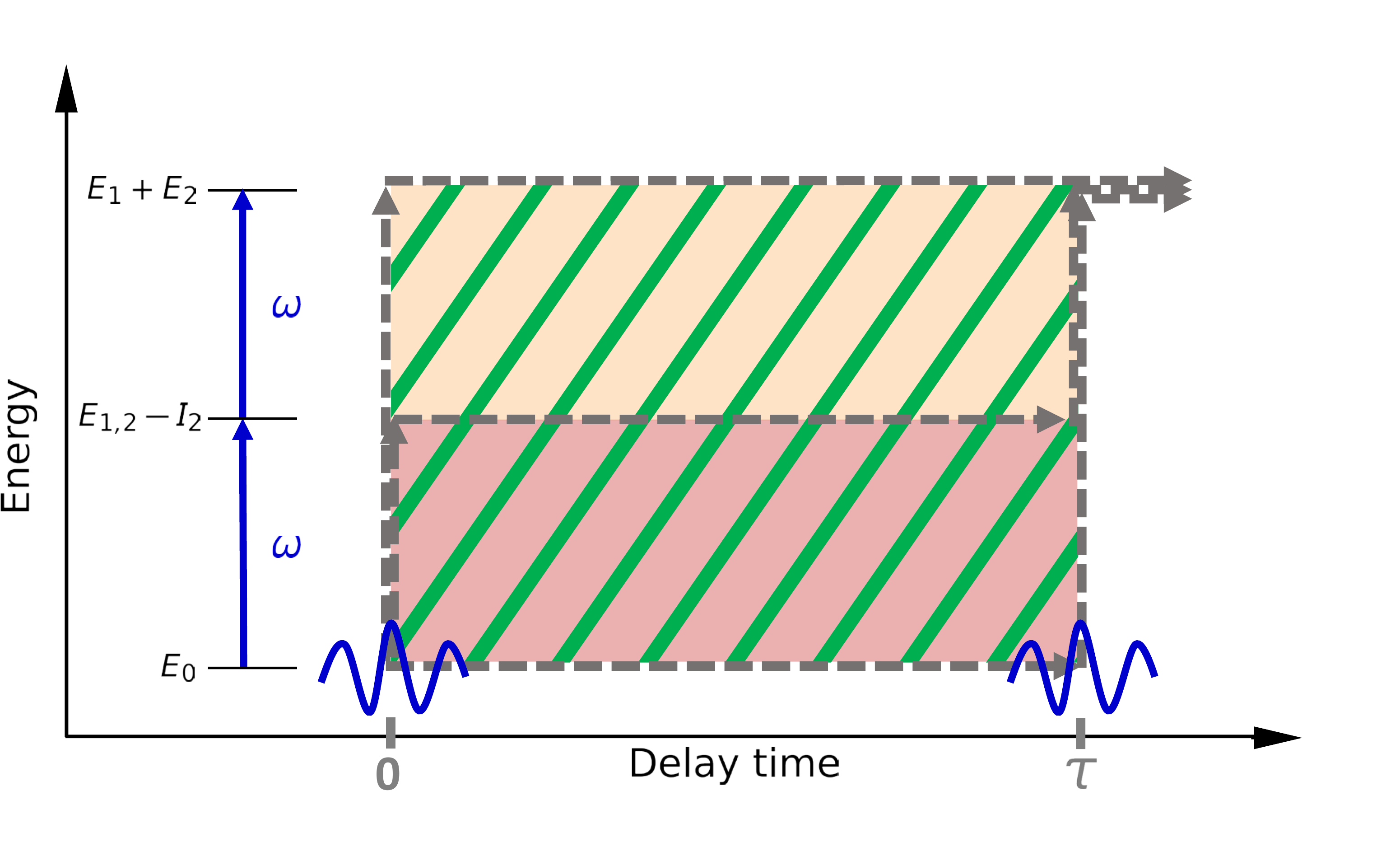} 
\caption{Energy-delay time diagram of the interfering pathways in a pump-probe sequence. 
The shaded areas signify the interference phases giving rise to the stripes (striped) and to the checkerboard pattern (red, orange shadow and striped).
For details see text.}\label{fig:paths}
\end{figure}
This pattern results from different pathways in the time domain to reach the same final two-electron state with energies ($E_1,E_2$) in the double continuum: absorption of two photons from the first pulse, absorption of two photons from the second pulse and absorption of one photon from each pulse, the latter amounting to two paths due to the indistinguishability of the two electrons (see Fig.~\ref{fig:paths}).
Closer inspection shows that the interference pattern near the sequential peaks is checker-board like while it appears stripe-like near the non-sequential ridge. \\
Employing second order perturbation theory (see, e.g., \cite{PazFeiNag2011,PalResMcc2009,HorMorRes2007,FouAntBac2008}) the interference pattern for two identical time-delayed XUV pulses can be  quantitatively accounted for.
The angle integrated two-photon double ionization probability is given by the incoherent sum of the transition probabilities $\left[t^{(2)}_{i \rightarrow f}\right]_{L,M}^2$  to all accessible $(L,M)$ channels \cite{PazFeiNag2011}
\begin{equation}
 P^{DI}\left(E_1,E_2\right) = \sum_{{L,M}=(2,2),(2,0),(2,-2),(0,0)} \left| \left[t^{(2)}_{i \rightarrow f}\right]_{L,M}\right|^2.
\end{equation}
For a pump-probe sequence where the first pulse has ellipticity $\epsilon_1$ and the second $\epsilon_2$ the amplitudes are given by
\begin{align} \label{eq:t_2_2}
\left[t^{(2)}_{i \rightarrow f}\right]_{2,\pm 2} \propto & \left(1 \pm \epsilon_1\right)^2 \mathcal{A}^{(2)} \left(\Delta E \right) e^{-i (E_1+E_2)\tau} \nonumber \\
&+ \left(1 \pm \epsilon_2\right)^2  \mathcal{A}^{(2)}  \left(\Delta E \right) e^{-i E_0\tau} \nonumber \\
& +\left(1 \pm \epsilon_1\right)\left(1 \pm \epsilon_2\right)  \mathcal{G}^{(1)}\left(\Delta E\right) e^{-i \left( E_1 -I_2\right)\tau}  \nonumber \\
& +\left(1 \pm \epsilon_1\right)\left(1 \pm \epsilon_2\right)  \mathcal{G}^{(1)}\left(-\Delta E\right) e^{-i \left( E_2 -I_2\right)\tau} , \\
\left[t^{(2)}_{i \rightarrow f}\right]_{2/0,0} \propto & \left( 1-\epsilon_{1}^2 \right)  \mathcal{A}^{(2)}  \left(\Delta E \right) e^{-i (E_1+E_2)\tau} \nonumber \\
& + \left( 1-\epsilon_{2}^2 \right)  \mathcal{A}^{(2)}  \left(\Delta E \right) e^{-i E_0\tau} \nonumber \\
& + \left(1-\epsilon_{1}\epsilon_{2}\right) \mathcal{G}^{(1)}\left(\Delta E\right) e^{-i \left( E_1 -I_2\right)\tau}  \nonumber \\
& + \left(1-\epsilon_{1}\epsilon_{2}\right)  \mathcal{G}^{(1)}\left(-\Delta E\right) e^{-i \left( E_2 -I_2\right)\tau}  \label{eq:t_0_0},   
\end{align}
where $\mathcal{A}^{(2)}$ is a generalized shape function for absorption of two photons within one pulse and $ \mathcal{G}^{(1)}$ is the shape function for absorbing one photon in each pulse (for details see Appendix~\ref{app:pattern}). 
$E_0=-\left(I_1+I_2\right)$ is the groundstate energy and $E_{1,2}-I_2$ is the intermediate state energy if the first (second) electron is ionized by the first pulse while the second (first) electron is bound in the 1s state of He$^{+}$.
Eq.~(\ref{eq:t_2_2}) represents the coherent superposition of ionization paths involving either absorption of two photons with the same polarization from the same pulse (terms proportional to $\mathcal{A}^{(2)}$) and absorption of photons from different pulses (terms proportional to $\mathcal{G}^{(1)}$) with, in general different polarizations $\epsilon_1 \neq \epsilon_2$.
The interference pattern depends on ($E_1,E_2$) or, equivalently, on $\Delta E=E_1-E_2$ and $E_{\mathrm{tot}}=E_1+E_2$.
Accordingly, the interference oscillations $\sim\cos\left(\left(E_1+E_2-E_0\right)\tau\right)$ along the lines $\Delta E$= const. are due to the mixed terms in $\mathcal{A}^{(2)}$ in Eq.~(\ref{eq:t_2_2}) and are, in principle, present in the entire ($E_1,E_2$) plane.
They appear, however, only when the circular polarization of the first and the second pulse agree.
The corresponding interference phase $\left(E_1+E_2-E_0\right)\tau$ is marked in Fig.~\ref{fig:paths} by green stripes.
Near the sequential peaks additional interference terms due to the $\mathcal{G}^{(1)}$ terms contribute (marked by red and yellow shaded areas in Fig.~\ref{fig:paths}) giving rise to oscillations $\sim\cos\left(\left(E_{1/2} - I_2 \right)\tau\right)$  $\left[\sim\cos\left(\left(E_{1/2}-I_1\right)\tau\right)\right]$  resulting from interferences between the third/fourth with the first (second) term in Eq.~(\ref{eq:t_2_2}) and, hence to the checkerboard pattern when the two polarizations agree. 
The third and fourth term do not interfere with each other since the spectral overlap between  $\mathcal{G}^{(1)} \left(\Delta E\right)$  and $\mathcal{G}^{(1)} \left(-\Delta E\right)$ vanishes. \\
Eq.~(\ref{eq:t_0_0}) contains terms that have significant weight only if the polarizations are sufficiently different and are strictly zero for $\epsilon_1=\epsilon_2=1$.
Thus by switching the polarizations of the pulses, path interferences can be switched on and off. \\
The interference in the double ionization spectrum  $P^{DI}(E_1,E_2)$ can be shown to be quite robust relative to variation of $\epsilon$ of the XUV pulses employed, as can be seen in Fig.~\ref{fig:pdi_e1_e2}~(c) for two pulses with $\epsilon_1=0.9$, $\epsilon_2=-0.9$.
The interference pattern starts to emerge, but is hardly visible.
Similarly, using two pulses with slightly different ellipticity, e.g., $\epsilon_1=0.9$, $\epsilon_2=0.6$ results in a very similar interference pattern as $\epsilon_1=\epsilon_2=1$ [compare Fig.~\ref{fig:pdi_e1_e2}~(b) and (d)]. \\
\subsection{Singly differential energy distributions}
\begin{figure}
\includegraphics[width=1\columnwidth]{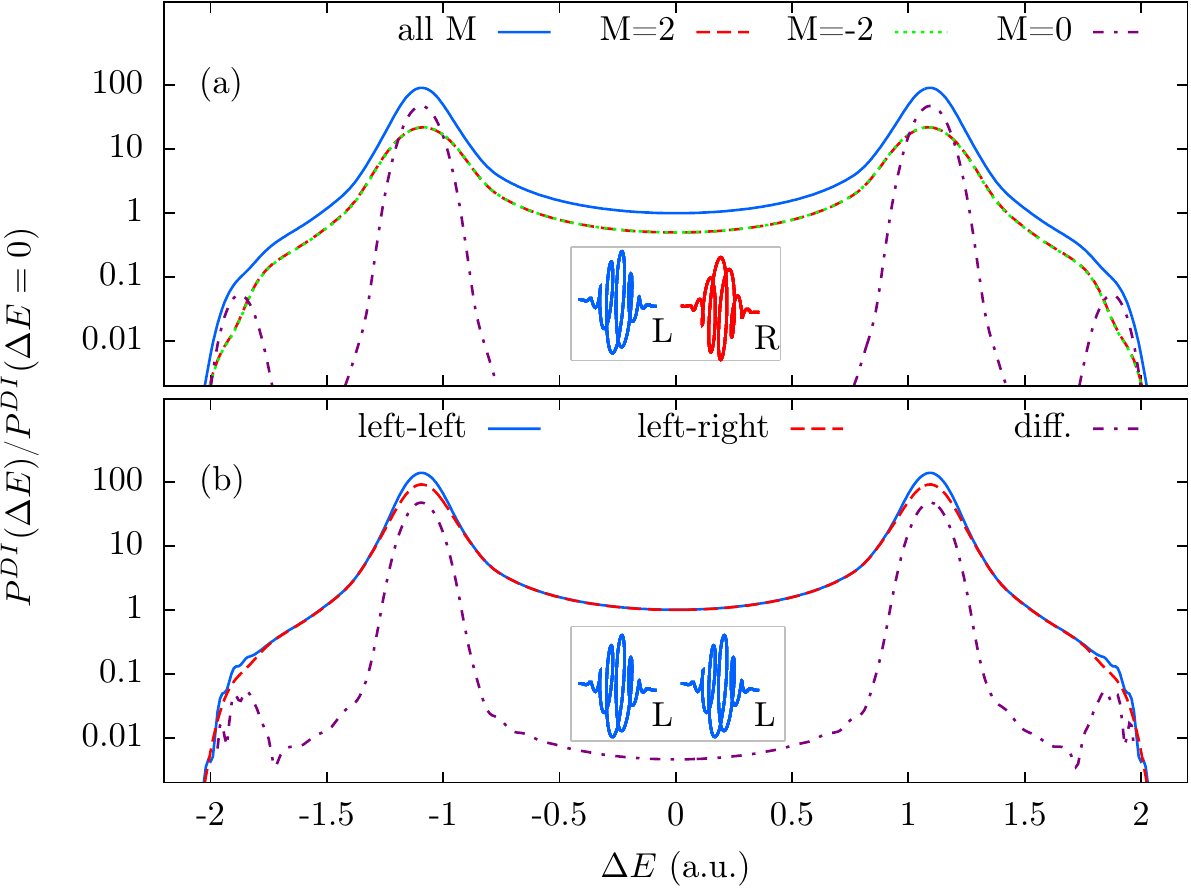}
\caption{
(a) $P^{DI}(\Delta E)$ for a pump-probe sequence with $\epsilon_1=$~1 ($L$), $\epsilon_2=$~$-$~1  ($R$) (see inset).
$P^{DI}(\Delta E)$ is resolved for the three different $M$ values of the final state.
(b) $P^{DI}(\Delta E)$  for a pump-probe sequence with two left-circularly polarized pulses ($\epsilon_1=\epsilon_2=$~1, see inset) and one left- and one right-circularly polarized pulse ($\epsilon_1=$~1, $\epsilon_2=$~$-$~1).
The dashed dotted line shows the difference of the two sequences.
The central photon energy of the pulses is 65 eV, their duration is $T_p$ = 1 fs and their delay is $\tau$= 2.5 fs.}\label{fig:pdi_delta_pump_probe}
\end{figure}

Integrating $P^{DI}(E_1,E_2)$ now along the total energy $E_{\mathrm{tot}}=E_1+E_2$  yields the double ionization probability as a function of the energy difference $\Delta E=E_1 - E_2$ of the two electrons
\begin{equation}\label{eq:pdi_delta_e}
P^{DI}(\Delta E)=\frac{1}{2}\int_0^{\infty} \text{d} E_{\mathrm{tot}} \:P^{DI}(E_1,E_2).
\end{equation}
As predicted by Eq.~(\ref{eq:t_2_2}) and (\ref{eq:t_0_0}), all the interference terms contain $\cos  \left( E_{\mathrm{tot}} \tau \right)$.
Integration over $E_{\mathrm{tot}}$ damps out all these terms.
Conversely, however, interferences due to the superposition of the terms $ \sim \mathcal{A}^{(2)}\sim \cos\left( \left( E_{\mathrm{tot}} -E_0 \right) \tau \right)$ survive when $P^{DI}\left(E_1,E_2\right)$ is integrated over $\Delta E$. \\
The singly-differential spectrum [Eq.~(\ref{eq:pdi_delta_e})] can be analyzed by making use of the polarization tagging.
This becomes possible because of the perfect locking of the magnetic quantum number of the two-electron wavepacket to the two ellipticities $\epsilon_{1,2}=1,-1$.
The electronic wavepacket with total $M$=2 (or $-$2) must have absorbed both photons from the first (second) pulse. 
$M$= 0 is reached only when one photon is absorbed from each pulse.
While the photo-electron spectrum of the $M=2$ $(-2)$ electrons is influenced by dynamical electron-electron correlation, the $M=0$ electrons stem from a purely sequential process.
This is clearly visible in $P^{DI}(\Delta E)$ [Fig.~\ref{fig:pdi_delta_pump_probe}~(a)].
Summing over all $M$ results in a $P^{DI}(\Delta E)$ closely resembling the distribution for a single linearly polarized XUV pulse \cite{PazFeiNag2011} with  a plateau in the equal energy sharing ($\Delta E\approx0$) region.
Analyzing $P^{DI}(\Delta E)$ separately for the different $M$ channels, we observe that for $M$=$\pm$2 the probability distribution is very similar to the full spectrum.
This is due to the fact that either channel consists of electrons emitted by the absorption of two photons within the same pulse and, thus, has the signatures of both temporal sequential and non-sequential two-photon double-ionization.
For $M$=0 [dash-dotted line in Fig.~\ref{fig:pdi_delta_pump_probe}~(a)],  $P^{DI}(\Delta E)$ features two well-separated peaks at $\Delta E \approx \pm 1.1$ a.u. while the equal-energy sharing plateau is completely missing as expected for a purely sequential ionization process.
Moreover, much smaller peaks near $\Delta E \approx \pm 1.9$ a.u. corresponding to shake-up processes with excited ionic states become visible. \\
The composition of TPDI into contributions with different total magnetic quantum number $M$ of the two-electron wavepacket is straightforward in theory, but such a separation is not easily accomplished in  experiments.
We therefore explore the signatures of polarization tagging in the experimentally directly accessible energy- and angular distributions of TPDI.
To this end, we compare the energy-differential double-ionization probability, $P^{DI}\left(\Delta E\right)$ for two polarization scenarios:  
A left-left $(L-L)$ pulse sequence ($\epsilon_1=\epsilon_2=$~1) with an $L-R$ sequence ($\epsilon_1=$~1, $\epsilon_2=$~$-$~1).
The absolute value of $P^{DI}\left(\Delta E=\pm 1.1 \: \mathrm{ a.u.} \right)$ near the sequential peak is found to be larger for the $L-L$ sequence than for the $L-R$ sequence.
By subtracting now $P^{DI}(\Delta E)$ of the $L-R$ sequence from $P^{DI}(\Delta E)$ of the $L-L$ sequence [dash-dotted line in Fig.~\ref{fig:pdi_delta_pump_probe}~(b)] the structure of the two well-separated sequential peaks emerges.
This difference spectrum coincides very well with the $M$=0 spectrum of the $L-R$ sequence [Fig.~\ref{fig:pdi_delta_pump_probe}~(a)] and gives access to the polarization tagged spectrum without actually resolving $M$.
This selectivity can be easily understood from angular momentum coupling arguments.
After absorbing the first left-handed photon, the singly ionized state lies in the $L$=1, $M$=1 channel.
During the second pulse the transition amplitude for absorbing another photon is the same irrespective of the polarization apart from the pre-factor which is given by a Clebsch-Gordon coefficient.
If the second pulse is also left-circularly polarized the only relevant Clebsch-Gordon coefficient for coupling to $M$=2 is $\langle 1, 1, 1, 1 | 2,2\rangle=1$.
For a right circularly-polarized second pulse there are two pathways to $M$=0 with Clebsch-Gordon coefficients $\langle 1, 1, 1, -1 | 2,0\rangle=1/\sqrt{6}$ and  $\langle 1, 1, 1, -1 | 0,0\rangle=1/\sqrt{3}$.
Squaring and adding those coefficients the probability to absorb one photon out of each pulse from an $L-R$ sequence is half of the $L-L$ sequence.
Consequently, by subtracting the $L-R$ from the $L-L$ signal the yield of the electrons is exactly the same as if we would have selected just the electrons which absorbed one photon out of each pulse in the $L-R$ scenario resulting in electrons with $M=0$. 
\subsection{Joint angular distributions}
\begin{figure}
\includegraphics[width=1\columnwidth]{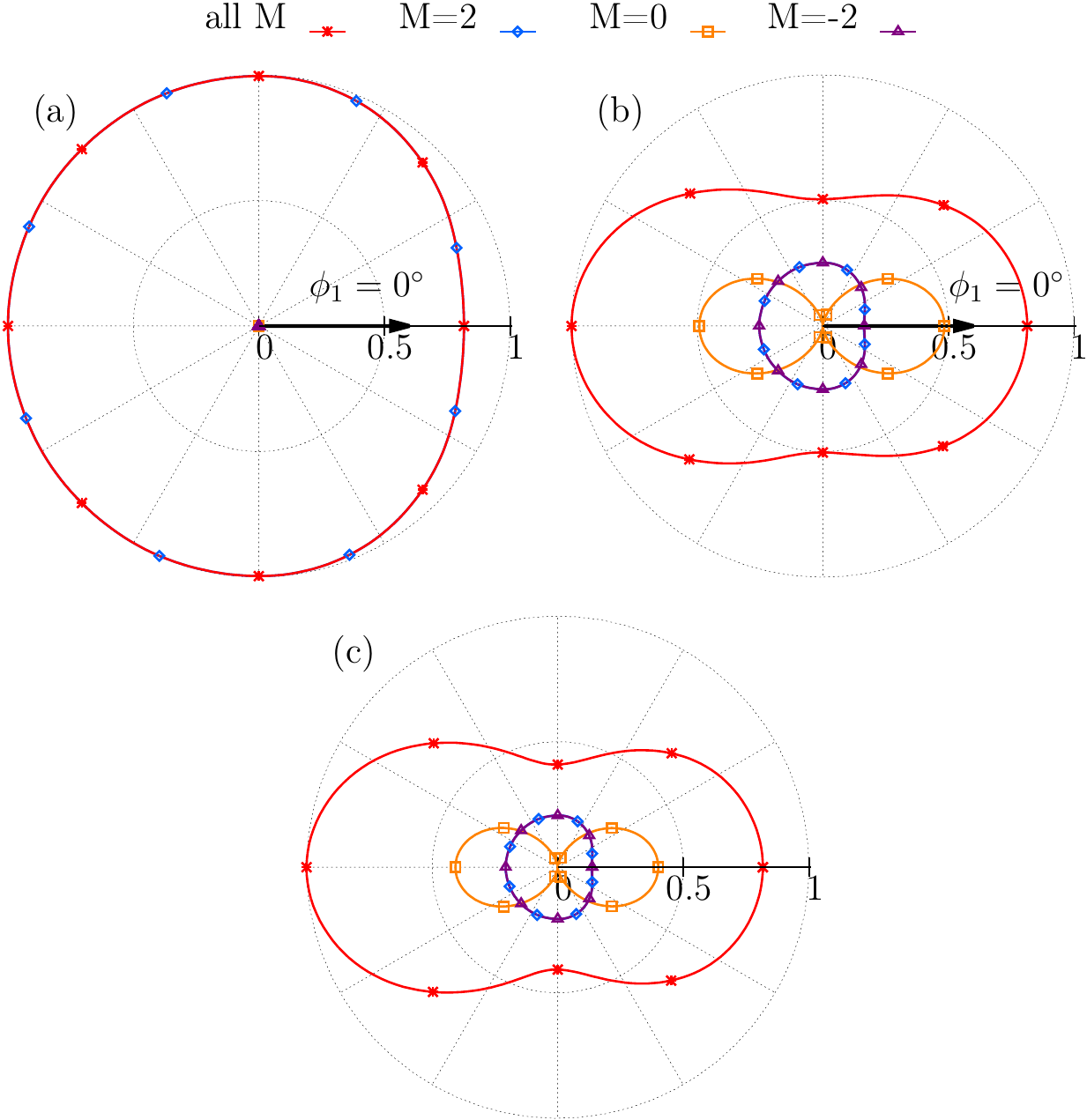}
\caption{
Joint angular distributions $P(\theta_1=\theta_2=$ 90$^{\circ},\phi_1=$ 0$^{\circ}, \phi_2)$ of the ejected electrons for an XUV pulse with polarization $\epsilon_1=1$ followed by an XUV pulse with (a) $\epsilon_2=1$ or (b) $\epsilon_2=-1$. 
(c) same as (b) for $\epsilon_1$=0.9  and $\epsilon_2$=$-$0.9.
All pulses have photon energy $\hbar \omega = $ 65 eV and a pulse duration of $T_p=$ 1 fs. 
The different lines correspond to different values of $M$ of the final state.
All distributions are normalized to the maximum value of the total probability (summed over all $M$).
 }\label{fig:tdcs_pump_probe}
\end{figure}

We now focus on the joint angular distributions of photoelectrons emitted in the XUV-XUV pulse sequence in the polarization plane ($\theta_1=\theta_2=90^{\circ}$) and fix the emission direction of the first electron to $\phi_1=0^{\circ}$.
We study differences and similarities in $P(\theta_1=\theta_2=$90$^{\circ},\phi_1=$0$^{\circ}, \phi_2)$ for the $L-L$ and $L-R$ pulse sequences [Fig.~\ref{fig:tdcs_pump_probe}~(a) and (b) respectively].
The angular distribution for an $L-L$ sequence [Fig.~\ref{fig:tdcs_pump_probe}~(a)] closely resembles $P(\phi_2)$ of a single pulse of comparable duration (Fig.~\ref{fig:tdcs_single_pulse_length}).
The distribution features a modest suppression due to dynamical Coulomb correlation for side-by-side emission but is otherwise close to rotationally symmetric.
Since the $L-L$ sequence has the propensity to populate $M$=2, the distribution from summation over all $M$ is essentially indistinguishable from the distribution for $M$=2 alone.
An entirely different picture emerges for the $L-R$ sequence [Fig.~\ref{fig:tdcs_pump_probe}~(b)].
The angular distribution $P(\phi_2)$ is peanut shaped.
Its origin can be traced to the contributions from different $M$ channels which contribute to $P(\phi_2)$ in the $L-R$ sequence.
While the $M =\pm$2 components mirror the distributions for the $L-L$ sequence, the $M$=0 component features a dipolar pattern with a nodal line along $\phi_2=$90$^{\circ}$ causing the indentation of $P(\phi_2)$.\\
This particular shape allows to associate the emission direction with the (non)sequential timing of the two-electron emission.
The joint probability near $P(\phi_2=$90$^{\circ})$ is exclusively due to the quasi-simultaneous emission from the same pulse.
Conversely, the joint probability near    $P(\phi_2=$0$^{\circ})$ is predominantly due to sequential emission of the two electrons spaced in time by $\tau$= 2.5 fs. 
In the present case of a relatively long pulse of $T_p$= 1 fs the sequential contribution is well above 50$\%$.
In the limit of even shorter pulses ($T_p \rightarrow 0$)  $P(\phi_2=$0$^{\circ})$ tends to become completely sequential [see Fig.~\ref{fig:tdcs_single_pulse_length}~(b)].
\\
To check for the robustness of our results against imperfection of circular polarization, we also perform simulations for a sequence with $\epsilon_1=0.9$,  $\epsilon_2=-0.9$.
The pulses have the same duration and energy as above.
While the overall yield decreases with decreasing ellipticity, the characteristic shape of the angular distribution remains largely unchanged and structurally stable [Fig.~\ref{fig:tdcs_pump_probe}~(c)].
It should be noted that the angular distribution for $\epsilon \neq \pm 1$ does not only depend on the relative angle $\phi_1-\phi_2$, but on both azimutal angles explicitly.
\section*{Conclusions}
We have explored the effect of ellipticity on two-photon double ionization, extending previous investigations of this two-electron process by linearly polarized fields, e.g. \cite{FeiNagPaz2008,LauBac2003,HuColCol2005,KheIva2006,ZhaPenXu2011,PalResMcc2009}.
We have shown that for a single ultrashort circularly polarized XUV pulse the imprint of dynamical Coulomb correlation on the joint angular distribution appears more prominently as the distortion due to the nodal structure of a dipolar emission pattern is absent.
For a sequence of two ultrashort XUV pulses with, in general, different polarizations it is possible by polarization tagging to control temporal interference patterns in the angle integrated double ionization probability $P^{DI}\left(E_1,E_2\right)$ and to map the timing of the emission events onto the joint angular distribution.
With new HHG based XUV sources with higher intensities as well as FEL based sources with improved control over the temporal structure of the light field these features should become experimentally accessible.   
\begin{acknowledgments}
The authors would like to thank Renate Pazourek and Fabian Lackner for helpful discussions. 
Calculations were performed on the Vienna Scientific Cluster (VSC3).
This work was supported by the WWTF through Project No. MA14-002, and  the  FWF  through  Projects  No.  FWF-SFB041-VICOM, No. FWF-SFB049-NEXTlite, No. FWF-W1243-Solids4Fun, as well as the IMPRS-APS.
\end{acknowledgments}

\appendix

\section{Interference pattern}\label{app:pattern}
Employing second-order time-dependent perturbation theory (see, e.g., \cite{PazFeiNag2011,PalResMcc2009,HorMorRes2007,FouAntBac2008}) the interference pattern observed in the angle-integrated double -ionization spectrum $P^{DI}\left(E_1,E_2\right)$ for a pump-probe sequence can be described quantitatively.
The transition amplitude for two-photon double ionization is accordingly given by
\begin{eqnarray}\label{eq:trans_ampl}
t^{(2)}_{i \rightarrow f}= &-  \:  \: \mathclap{\displaystyle\int_n}\mathclap{\textstyle\sum} \: \: \int_{-\infty}^{t_f} \mathrm{d} t_1 \int_{-\infty}^{t_1} \mathrm{d} t_2 e^{iE_{fn}t_1} e^{iE_{ni}t_2} \nonumber \\
& \times \left< f \left| V(t_1) \right| n\right>  \left< n \left| V(t_2) \right|  i \right>,
\end{eqnarray}
with $E_{fn}=E_f-E_n$ and  $E_{ni}=E_n-E_i$.
The initial state $\left|i\right>$ is the ground state and the final state $\left|f\right>=\left| E_1 \Omega_1, E_2 \Omega_2 \right>$ is the double continuum.
The sum runs over all intermediate two-electron states $\left|n\right>$. \\
To calculate the transition amplitude to the final state for the different $(L,M)$ channels $\left[t^{(2)}_{i \rightarrow f}\right]_{L,M}$ we project onto the final state $\left|E_1,E_2,L,M\right>$.
The two-photon double ionization probability is then given by
\begin{equation}
 P^{DI}\left(E_1,E_2\right) = \sum_{{L,M}=(2,2),(2,0),(2,-2),(0,0)} \left| \left[t^{(2)}_{i \rightarrow f}\right]_{L,M}\right|^2.
\end{equation}
We employ the velocity gauge, i.e. $\hat{V}(t)=\left(\hat{\vec{p}}_1+\hat{\vec{p}}_2\right)\vec{A}(t)\equiv\hat{\vec{\mu}}\vec{A}(t)$.
Expanding $\hat{\vec{\mu}}\vec{A}(t)$ in spherical tensor components yields 
\begin{equation}\label{eq:spherical_tensors}
\hat{\vec{\mu}}\vec{A}(t)=\sum_{q=-1}^{1} \left(-1\right)^q \hat{\mu}_q A_{-q}, 
\end{equation}
where the spherical tensor components are defined as $\hat{T}_{+1}= - \left( \hat{T}_x + i \hat{T}_y \right)/\sqrt{2}$, $\hat{T}_{-1}=\left(\hat{T}_x- i \hat{T}_y\right)/\sqrt{2}$, $\hat{T}_{0}=\hat{T}_z$.
Using the vector potential [Eq.~(\ref{eq:field})] in rotating-wave approximation $\left[\vec{A}(t) \approx A_0/2 \: f(t)  e^{-i \omega t}  \left(i \hat{x} - \epsilon \hat{y} \right)\right]$ the spherical components of the field are $A_{+1}=-i\left(1-\epsilon\right) A_0 f(t)  e^{-i \omega t} /\sqrt{8}$, $A_{-1}=i \left(1+\epsilon\right)A_0 f(t)  e^{-i \omega t} /\sqrt{8}$, $A_{0}=0$.
Inserting into Eq.~(\ref{eq:spherical_tensors}) yields
\begin{align}\label{eq:interaction_op}
\hat{\vec{\mu}}\vec{A}(t)&= \left[ - \hat{\mu}_{+1}  \left(1+\epsilon\right)  + \hat{\mu}_{-1}  \left(1-\epsilon \right) \right] i \frac{A_0}{\sqrt{8}} f(t)  e^{-i \omega t} \nonumber \\
& = \hat{\mu}_{\epsilon} \bar{A}(t).
\end{align}
Within the rotating-wave approximation $\hat{\mu}_{\epsilon}$ can be expanded as a function of spherical harmonics
\begin{equation}
\hat{\mu}_{\epsilon} \propto -\left(1+\epsilon \right) Y_1^{1} + \left(1-\epsilon \right) Y_1^{-1}.
\end{equation} 
Using Eq.~(\ref{eq:interaction_op}) we can simplify Eq.~(\ref{eq:trans_ampl}):
\begin{align}\label{eq:trans_ampl_l_m}
\left[t^{(2)}_{i \rightarrow f}\right]_{L,M}=-\:  \: \: \mathclap{\displaystyle\int_n}\mathclap{\textstyle\sum} \: \: \: & \left< E_1,E_2,L,M \left| \hat{\mu}_{\epsilon}\right| n\right>  \left< n \left|\hat{\mu}_{\epsilon} \right|  i \right> \nonumber\\
& \times  \: \mathcal{G} \left[ E_{fn},E_{ni},\bar{A}(t)  \right],
\end{align}
with
\begin{align}\label{eq:shape_function_gen}
\mathcal{G}& \left[ E_{fn},E_{ni},\bar{A}(t)  \right]= \nonumber \\
&\int_{-\infty}^{t_f} \mathrm{d} t_1 \int_{-\infty}^{t_1} \mathrm{d} t_2 e^{iE_{fn}t_1} e^{iE_{ni}t_2} \bar{A}(t_1) \bar{A}(t_2).
\end{align}
We note that the shape function $\mathcal{G}$ is fully determined by the temporal envelope $f(t)$ and the central frequency of the pulse.
To leading order only the intermediate state with one electron in the $p$ continuum and the other electron bound in the $1s$ state of He$^{+}$ contributes ($\left|n_0\right>=\left|E_1 p, 1s\right>$).
The contribution of shake-up intermediate states is small (see Fig.~\ref{fig:pdi_delta_pump_probe}). \\
Assuming only the on-shell intermediate state by absorbing the first photon with ellipticity  $\epsilon_{I}$ and the second with ellipticity $\epsilon_{II}$ the matrix element in Eq.~\ref{eq:trans_ampl_l_m} is given by 
\begin{align}
M_{2,2}= &- \left(1+\epsilon_{II}\right) \left( 1+\epsilon_{I}\right) C_{2,2;1,1} \nonumber \\
M_{2,0}= &  \left(1+\epsilon_{II}\right)  \left(1-\epsilon_{I}\right) C_{2,0;1,-1}  \nonumber \\
&+  \left(1-\epsilon_{II}\right) \left(1+\epsilon_{I}\right) C_{2,0;1,1} \nonumber \\
=&  2\left(1-\epsilon_{I} \epsilon_{II}\right)C_{2,0;1,1} \nonumber \\
M_{2,-2}= & - \left(1-\epsilon_{II}\right) \left( 1-\epsilon_{I}\right) C_{2,2;1,-1} \nonumber \\
M_{0,0}= &  \left(1+\epsilon_{II}\right)  \left(1-\epsilon_{I}\right) C_{0,0;1,-1}  \nonumber \\
&+  \left(1-\epsilon_{II}\right) \left(1+\epsilon_{I}\right) C_{0,0;1,1} \nonumber \\ 
=&  2\left(1-\epsilon_{I} \epsilon_{II}\right)C_{0,0;1,1} ,
\end{align}
with $C_{L,M;L',M'}$ being the product of Clebsch-Gordon coefficients and the reduced matrix elements containing the momentum operators $\hat{p}_{+1}$ and $\hat{p}_{-1}$. 
$(L',M')$ are the total angular momentum and magnetic quantum number of the intermediate state $\left| n_0 \right>$ and $ C_{2,0;1,-1}= C_{2,0;1,+1}$, $ C_{0,0;1,-1}= C_{0,0;1,+1}$. \\
To calculate the shape function $\mathcal{G}$ [Eq.~(\ref{eq:shape_function_gen})] we distinguish whether both photons are absorbed within one pulse [$\mathcal{G}^{(2)}$] or one photon out of each pulse is absorbed  [$\mathcal{G}^{(1)}$].
$\mathcal{G}^{(1)}$ can be easily calculated since the integration limits in Eq.~(\ref{eq:shape_function_gen}) are independent
\begin{align}
&\mathcal{G}^{(1)}  \left[ E_{fn},E_{ni},\bar{A}(t)  \right]= \nonumber \\& \int_{-\infty}^{t_f} \mathrm{d} t_1  e^{iE_{fn}t_1}  \bar{A}\left(t_1 \right) \int_{-\infty}^{t_f} \mathrm{d} t_2e^{iE_{ni}t_2} \bar{A}\left(t_2 \right).
\end{align}
Inserting the envelope of the pulse we get
\begin{align}
& \mathcal{G}^{(1)} \left( E_{fn},  E_{ni},A_0,T_p  \right) =  \frac{A_0^2}{8} \nonumber \\
& \int_{-T_p}^{T_p} \mathrm{d} t_1 e^{i  \left( E_{ni} -\omega \right)  t_1 } \cos^2\left(\frac{\pi t_1}{2 T_p}\right)  \nonumber \\
& \int_{-T_p}^{T_p} \mathrm{d} t_2 e^{i \left( E_{fn} -\omega \right) t_2} \cos^2\left(\frac{\pi t_2}{2 T_p}\right).
\end{align} 
For long pulses the total energy of the final state is accurately determined to be $E_1+E_2=2\omega+E_0$ and we can write $\mathcal{G}^{(1)}$ as function of $(E_{\mathrm{tot}},\Delta E)$ 
\begin{align}
& \mathcal{G}^{(1)} \left( \Delta E \right) =  \frac{A_0^2}{8} \nonumber \\
& \int_{-T_p}^{T_p} \mathrm{d} t_1 e^{i  \left( \Delta E +I_1-I_2 \right)  t_1 /2 } \cos^2\left(\frac{\pi t_1}{2 T_p}\right)  \nonumber \\
& \int_{-T_p}^{T_p} \mathrm{d} t_2 e^{i \left( -\Delta E +I_2-I_1 \right)  t_2 /2} \cos^2\left(\frac{\pi t_2}{2 T_p}\right).
\end{align} 
$\mathcal{G}^{(1)} \left( \Delta E \right)$ depends only on $\Delta E$ but not on $E_{\mathrm{tot}}$. 
Furthermore, for sufficiently long pulses $\mathcal{G}^{(1)} \left( \Delta E \right)$  and  $\mathcal{G}^{(1)} \left( - \Delta E \right)$  appearing in Eqs.~(\ref{eq:t_2_2}),(\ref{eq:t_0_0}) have vanishing spectral overlap.
Consequently mixed terms of the form $\mathcal{G}^{(1)} \left( \Delta E \right) \left[   \mathcal{G}^{(1)} \left( -\Delta E \right) \right]^{*}$ vanish.  \\
$\mathcal{G}^{(2)}$  is given by  \cite{PazFeiNag2011} 
\begin{align}
& \mathcal{G}^{(2)} \left( \Delta E_{fn}, \Delta E_{ni},A_0,T_p  \right) \nonumber \\
& = \frac{A_0^2}{8}\int_{-T_p}^{ T_p} \mathrm{d} t_1 \int_{-T_p}^{t_1} \mathrm{d} t_2 \mathcal{F} \left( \Delta E_{fn}, t_1, T_p\right)  \mathcal{F}\left( \Delta E_{ni}, t_2, T_p\right),
\end{align} 
with
\begin{equation}
\mathcal{F}\left( \xi, t, T\right) = e^{i \xi t} \cos^2\left(\frac{\pi t}{2T}\right),
\end{equation}
where $\Delta E_{fn}=E_{fn}-\omega$ and $\Delta E_{ni}=E_{ni}-\omega$.
As for $\mathcal{G}^{(1)}$ we use that for long pulses the total energy of the final state is accurately determined to be $E_1+E_2=2\omega+E_0$.
This enforces $\Delta E_{fn}=-\Delta E_{ni}$.
Changing the variables from $(E_1,E_2)$ to $(E_{\mathrm{tot}},\Delta E)$ one obtains
\begin{align}
\mathcal{G}^{(2)} & \left( \Delta E_{fn}, \Delta E_{ni},A_0,T_p  \right)  =  \nonumber \\
& \mathcal{G}^{(2)} \left(  \left(I_2-I_1-\Delta E\right)/2,-\left(I_2-I_1-\Delta E\right)/2,A_0,T_p  \right). 
\end{align}
Not only the shape function $\mathcal{G}^{(1)}$ but also  $\mathcal{G}^{(2)}$ depend only on $\Delta E$ and not on $E_{\mathrm{tot}}$.
As shown in \cite{PazFeiNag2011} the amplitude of two-photon absorption within one pulse is given by $\mathcal{A}^{(2)}=\mathcal{G}^{(2)}\left(\Delta E\right) + \mathcal{G}^{(2)}\left(-\Delta E\right)$. \\
For a pump-probe sequence where the first pulse has ellipticity $\epsilon_1$ and the second $\epsilon_2$ the transition amplitudes are given by
\begin{align}
\left[t^{(2)}_{i \rightarrow f}\right]_{2,2} \propto & \left(1+\epsilon_1\right)^2 \mathcal{A}^{(2)} \left(\Delta E\right) e^{-i (E_1+E_2)\tau} \nonumber \\
&+ \left(1+\epsilon_2\right)^2  \mathcal{A}^{(2)}\left(\Delta E\right) e^{-i E_0\tau} \nonumber \\
& +\left(1+ \epsilon_1\right)\left(1+ \epsilon_2\right) \mathcal{G}^{(1)}\left(\Delta E\right) e^{-i \left( E_1 -I_2\right)\tau}  \nonumber \\
& +\left(1+ \epsilon_1\right)\left(1+ \epsilon_2\right)   \mathcal{G}^{(1)}\left(-\Delta E\right) e^{-i \left( E_2 -I_2\right)\tau},
\end{align}
\begin{align}
\left[t^{(2)}_{i \rightarrow f}\right]_{2,-2} \propto & \left(1-\epsilon_1\right)^2 \mathcal{A}^{(2)} \left(\Delta E\right) e^{-i  (E_1+E_2)\tau} \nonumber \\
& + \left(1-\epsilon_2\right)^2  \mathcal{A}^{(2)}\left(\Delta E\right)  e^{-i E_0\tau} \nonumber \\
& +\left(1- \epsilon_1\right)\left(1- \epsilon_2\right) \mathcal{G}^{(1)}\left(\Delta E\right) e^{-i \left( E_1 -I_2\right)\tau}  \nonumber \\
& + \left(1- \epsilon_1\right)\left(1- \epsilon_2\right)  \mathcal{G}^{(1)}\left(-\Delta E\right) e^{-i \left( E_2 -I_2\right)\tau},
\end{align}
\begin{align}
\left[t^{(2)}_{i \rightarrow f}\right]_{2,0} \propto & \left( 1-\epsilon_{1}^2 \right)  \mathcal{A}^{(2)}\left(\Delta E\right) e^{-i  (E_1+E_2)\tau} \nonumber \\
& + \left( 1-\epsilon_{2}^2 \right)  \mathcal{A}^{(2)} \left(\Delta E\right) e^{-i E_0\tau} \nonumber \\
& + \left(1-\epsilon_{1}\epsilon_{2}\right) \mathcal{G}^{(1)}\left(\Delta E\right) e^{-i \left( E_1 -I_2\right)\tau} \nonumber \\
& +  \left(1-\epsilon_{1}\epsilon_{2}\right) \mathcal{G}^{(1)}\left(-\Delta E\right) e^{-i \left( E_2 -I_2\right)\tau},      
\end{align}
\begin{align}
\left[t^{(2)}_{i \rightarrow f}\right]_{0,0} \propto & \left( 1-\epsilon_{1}^2 \right)  \mathcal{A}^{(2)} \left(\Delta E\right)  e^{-i (E_1+E_2)\tau} \nonumber \\
& + \left( 1-\epsilon_{2}^2 \right)  \mathcal{A}^{(2)}\left(\Delta E\right) e^{-i E_0\tau} \nonumber \\
&  + \left(1-\epsilon_{1}\epsilon_{2}\right) \mathcal{G}^{(1)}\left(\Delta E\right) e^{-i \left( E_1 -I_2\right)\tau}  \nonumber \\ 
&  +  \left(1-\epsilon_{1}\epsilon_{2}\right)  \mathcal{G}^{(1)}\left(-\Delta E\right) e^{-i \left( E_2 -I_2\right)\tau} .
\end{align}
For the $L-L$ and the $L-R$ sequence these equations simplify drastically .  \\
For the $L-L$ sequence
\begin{align}
\left[t^{(2)}_{i \rightarrow f}\right]_{2,2} \propto &  \mathcal{A}^{(2)} \left(\Delta E\right) \left( e^{-i  (E_1+E_2)\tau}+  e^{-i E_0\tau}  \right)   \nonumber \\
& +  \mathcal{G}^{(1)}\left(\Delta E\right) e^{-i \left( E_1 -I_2\right)\tau}  \nonumber \\
&+  \mathcal{G}^{(1)}\left(-\Delta E\right) e^{-i \left( E_2 -I_2\right)\tau} \nonumber 
\end{align}
while
\begin{align}
\left[t^{(2)}_{i \rightarrow f}\right]_{2,0} &= \left[t^{(2)}_{i \rightarrow f}\right]_{2,-2} =\left[t^{(2)}_{i \rightarrow f}\right]_{0,0} =   0 \nonumber.
\end{align}
Consequently the $\left[t^{(2)}_{i \rightarrow f}\right]_{2,2}$ terms fully account for the interferences.
For the $L-R$ sequence we have:
\begin{align}
\left[t^{(2)}_{i \rightarrow f}\right]_{2,2} \propto &  \mathcal{A}^{(2)} \left(\Delta E\right) e^{-i  (E_1+E_2)\tau} , \nonumber 
\end{align}
\begin{align}
\left[t^{(2)}_{i \rightarrow f}\right]_{2,-2} \propto &   \mathcal{A}^{(2)} \left(\Delta E\right) e^{-i E_0\tau},  \nonumber 
\end{align}
\begin{align}
\left[t^{(2)}_{i \rightarrow f}\right]_{2,0} \propto &   \mathcal{G}^{(1)}\left(\Delta E\right) e^{-i \left( E_1 -I_2\right)\tau}  \nonumber  \\
&  +  \mathcal{G}^{(1)}\left(-\Delta E\right) e^{-i \left( E_2 -I_2\right)\tau} ,    \nonumber 
\end{align}
\begin{align}
\left[t^{(2)}_{i \rightarrow f}\right]_{0,0} \propto &  \mathcal{G}^{(1)}\left(\Delta E\right) e^{-i \left( E_1 -I_2\right)\tau} \nonumber \\
&  +  \mathcal{G}^{(1)}\left(-\Delta E\right) e^{-i \left( E_2 -I_2\right)\tau}  . \nonumber
\end{align}
Since  $\mathcal{G}^{(1)}(\Delta E)$ and $\mathcal{G}^{(1)}(-\Delta E)$ have vanishing spectral overlap the only interfering terms in $\left[t^{(2)}_{i \rightarrow f}\right]_{2,0}$ and $\left[t^{(2)}_{i \rightarrow f}\right]_{0,0}~\sim \mathcal{G}^{(1)} \left( \Delta E \right) \left[   \mathcal{G}^{(1)} \left( -\Delta E \right) \right]^{*}$ vanish.
Consequently there are effectively no interfering paths and hence no interference pattern in $P^{DI}\left(E_1,E_2\right)$ for the $L-R$ sequence.


\begin{thebibliography}{10}

\bibitem{KraSchKul1992}
J.~L. Krause, K.~J. Schafer, and K.~C. Kulander.
\newblock \emph{High-order harmonic generation from atoms and ions in the high
  intensity regime}.
\newblock Physical Review Letters \textbf{68}, 3535 (1992).

\bibitem{MacKmeGor1993}
J.~J. Macklin, J.~D. Kmetec, and C.~L. Gordon.
\newblock \emph{High-order harmonic generation using intense femtosecond
  pulses}.
\newblock Physical Review Letters \textbf{70}, 766 (1993).

\bibitem{PopCheAr2010}
T.~Popmintchev, M.-C. Chen, P.~Arpin, M.~M. Murnane, and H.~C. Kapteyn.
\newblock \emph{The attosecond nonlinear optics of bright coherent {X}-ray
  generation}.
\newblock Nature Photonics \textbf{4}, 822 (2010).

\bibitem{FleKfuDis2014}
A.~Fleischer, O.~Kfir, T.~Diskin, P.~Sidorenko, and O.~Cohen.
\newblock \emph{Spin angular momentum and tunable polarization in high-harmonic
  generation}.
\newblock Nature Photonics \textbf{8}, 543 (2014).

\bibitem{Mil2015}
D.~B. Milošević.
\newblock \emph{Circularly polarized high harmonics generated by a bicircular
  field from inert atomic gases in the p state: {A} tool for exploring
  chirality-sensitive processes}.
\newblock Physical Review A \textbf{92}, 043827 (2015).

\bibitem{BakAhsLin2016}
D.~Baykusheva, M.~S. Ahsan, N.~Lin, and H.~J. Wörner.
\newblock \emph{Bicircular {High}-{Harmonic} {Spectroscopy} {Reveals}
  {Dynamical} {Symmetries} of {Atoms} and {Molecules}}.
\newblock Physical Review Letters \textbf{116}, 123001 (2016).

\bibitem{MauBanUze2016}
F.~Mauger, A.~D. Bandrauk, and T.~Uzer.
\newblock \emph{Circularly polarized molecular high harmonic generation using a
  bicircular laser}.
\newblock Journal of Physics B: Atomic, Molecular and Optical Physics
  \textbf{49}, 10LT01 (2016).

\bibitem{ReiMad2016}
D.~M. Reich and L.~B. Madsen.
\newblock \emph{Illuminating {Molecular} {Symmetries} with {Bicircular}
  {High}-{Order}-{Harmonic} {Generation}}.
\newblock Physical Review Letters \textbf{117}, 133902 (2016).

\bibitem{MedWraHar2015}
L.~Medišauskas, J.~Wragg, H.~van~der Hart, and M.~Y. Ivanov.
\newblock \emph{Generating {Isolated} {Elliptically} {Polarized} {Attosecond}
  {Pulses} {Using} {Bichromatic} {Counterrotating} {Circularly} {Polarized}
  {Laser} {Fields}}.
\newblock Physical Review Letters \textbf{115}, 153001 (2015).

\bibitem{KfiGryTur2015}
O.~Kfir, P.~Grychtol, E.~Turgut, R.~Knut, D.~Zusin, D.~Popmintchev,
  T.~Popmintchev, H.~Nembach, J.~M. Shaw, A.~Fleischer, H.~Kapteyn, M.~Murnane,
  and O.~Cohen.
\newblock \emph{Generation of bright phase-matched circularly-polarized extreme
  ultraviolet high harmonics}.
\newblock Nature Photonics \textbf{9}, 99 (2015).

\bibitem{FanGryKnu2015}
T.~Fan, P.~Grychtol, R.~Knut, C.~Hernández-García, D.~D. Hickstein, D.~Zusin,
  C.~Gentry, F.~J. Dollar, C.~A. Mancuso, C.~W. Hogle, O.~Kfir, D.~Legut,
  K.~Carva, J.~L. Ellis, K.~M. Dorney, C.~Chen, O.~G. Shpyrko, E.~E. Fullerton,
  O.~Cohen, P.~M. Oppeneer, D.~B. Milošević, A.~Becker, A.~A. Jaroń-Becker,
  T.~Popmintchev, M.~M. Murnane, and H.~C. Kapteyn.
\newblock \emph{Bright circularly polarized soft {X}-ray high harmonics for
  {X}-ray magnetic circular dichroism}.
\newblock Proceedings of the National Academy of Sciences \textbf{112}, 14206
  (2015).

\bibitem{CheTaoHer2016}
C.~Chen, Z.~Tao, C.~Hernández-García, P.~Matyba, A.~Carr, R.~Knut, O.~Kfir,
  D.~Zusin, C.~Gentry, P.~Grychtol, O.~Cohen, L.~Plaja, A.~Becker,
  A.~Jaron-Becker, H.~Kapteyn, and M.~Murnane.
\newblock \emph{Tomographic reconstruction of circularly polarized
  high-harmonic fields: 3D attosecond metrology}.
\newblock Science Advances \textbf{2}, e1501333 (2016).

\bibitem{HuaHerHua2018}
P.-C. Huang, C.~Hernández-García, J.-T. Huang, P.-Y. Huang, C.-H. Lu,
  L.~Rego, D.~D. Hickstein, J.~L. Ellis, A.~Jaron-Becker, A.~Becker, S.-D.
  Yang, C.~G. Durfee, L.~Plaja, H.~C. Kapteyn, M.~M. Murnane, A.~H. Kung, and
  M.-C. Chen.
\newblock \emph{Polarization control of isolated high-harmonic pulses}.
\newblock Nature Photonics \textbf{12}, 349 (2018).

\bibitem{ChePuk2016}
Z.-Y. Chen and A.~Pukhov.
\newblock \emph{Bright high-order harmonic generation with controllable
  polarization from a relativistic plasma mirror}.
\newblock Nature Communications \textbf{7}, 12515 (2016).

\bibitem{MaYuYu2016}
G.~Ma, W.~Yu, M.~Y. Yu, B.~Shen, and L.~Veisz.
\newblock \emph{Intense circularly polarized attosecond pulse generation from
  relativistic laser plasmas using few-cycle laser pulses}.
\newblock Optics Express \textbf{24}, 10057 (2016).

\bibitem{AllDivCal2014}
E.~Allaria, B.~Diviacco, C.~Callegari, P.~Finetti, B.~Mahieu, J.~Viefhaus,
  M.~Zangrando, G.~De~Ninno, G.~Lambert, E.~Ferrari, J.~Buck, M.~Ilchen,
  B.~Vodungbo, N.~Mahne, C.~Svetina, C.~Spezzani, S.~Di~Mitri, G.~Penco,
  M.~Trovó, W.~M. Fawley, P.~R. Rebernik, D.~Gauthier, C.~Grazioli, M.~Coreno,
  B.~Ressel, A.~Kivimäki, T.~Mazza, L.~Glaser, F.~Scholz, J.~Seltmann,
  P.~Gessler, J.~Grünert, A.~De~Fanis, M.~Meyer, A.~Knie, S.~P. Moeller,
  L.~Raimondi, F.~Capotondi, E.~Pedersoli, O.~Plekan, M.~B. Danailov,
  A.~Demidovich, I.~Nikolov, A.~Abrami, J.~Gautier, J.~Lüning, P.~Zeitoun, and
  L.~Giannessi.
\newblock \emph{Control of the {Polarization} of a {Vacuum}-{Ultraviolet},
  {High}-{Gain}, {Free}-{Electron} {Laser}}.
\newblock Physical Review X \textbf{4}, 041040 (2014).

\bibitem{LutMaxIlc2016}
A.~A. Lutman, J.~P. MacArthur, M.~Ilchen, A.~O. Lindahl, J.~Buck, R.~N. Coffee,
  G.~L. Dakovski, L.~Dammann, Y.~Ding, H.~A. Dürr, L.~Glaser, J.~Grünert,
  G.~Hartmann, N.~Hartmann, D.~Higley, K.~Hirsch, Y.~I. Levashov, A.~Marinelli,
  T.~Maxwell, A.~Mitra, S.~Moeller, T.~Osipov, F.~Peters, M.~Planas,
  I.~Shevchuk, W.~F. Schlotter, F.~Scholz, J.~Seltmann, J.~Viefhaus, P.~Walter,
  Z.~R. Wolf, Z.~Huang, and H.-D. Nuhn.
\newblock \emph{Polarization control in an {X}-ray free-electron laser}.
\newblock Nature Photonics \textbf{10}, 468 (2016).

\bibitem{KorWedNol2017}
C.~von Korff~Schmising, D.~Weder, T.~Noll, B.~Pfau, M.~Hennecke, C.~Strüber,
  I.~Radu, M.~Schneider, S.~Staeck, C.~M. Günther, J.~Lüning, A.~e.~d. Merhe,
  J.~Buck, G.~Hartmann, J.~Viefhaus, R.~Treusch, and S.~Eisebitt.
\newblock \emph{Generating circularly polarized radiation in the extreme
  ultraviolet spectral range at the free-electron laser {FLASH}}.
\newblock Review of Scientific Instruments \textbf{88}, 053903 (2017).

\bibitem{NgoHuMad2015}
J.~Ngoko~Djiokap, S.~Hu, L.~Madsen, N.~Manakov, A.~Meremianin, and A.~F.
  Starace.
\newblock \emph{Electron {Vortices} in {Photoionization} by {Circularly}
  {Polarized} {Attosecond} {Pulses}}.
\newblock Physical Review Letters \textbf{115}, 113004 (2015).

\bibitem{PenKerJoh2017}
D.~Pengel, S.~Kerbstadt, D.~Johannmeyer, L.~Englert, T.~Bayer, and
  M.~Wollenhaupt.
\newblock \emph{Electron {Vortices} in {Femtosecond} {Multiphoton}
  {Ionization}}.
\newblock Physical Review Letters \textbf{118}, 053003 (2017).

\bibitem{FeiNagPaz2008}
J.~Feist, S.~Nagele, R.~Pazourek, E.~Persson, B.~I. Schneider, L.~A. Collins,
  and J.~Burgdörfer.
\newblock \emph{Nonsequential two-photon double ionization of helium}.
\newblock Physical Review A \textbf{77}, 043420 (2008).

\bibitem{FeiNagPaz2009}
J.~Feist, S.~Nagele, R.~Pazourek, E.~Persson, B.~I. Schneider, L.~A. Collins,
  and J.~Burgdörfer.
\newblock \emph{Probing {Electron} {Correlation} via {Attosecond} xuv {Pulses}
  in the {Two}-{Photon} {Double} {Ionization} of {Helium}}.
\newblock Physical Review Letters \textbf{103}, 063002 (2009).

\bibitem{PazFeiNag2011}
R.~Pazourek, J.~Feist, S.~Nagele, E.~Persson, B.~I. Schneider, L.~A. Collins,
  and J.~Burgdörfer.
\newblock \emph{Universal features in sequential and nonsequential two-photon
  double ionization of helium}.
\newblock Physical Review A \textbf{83}, 053418 (2011).

\bibitem{FeiNagTic2011}
J.~Feist, S.~Nagele, C.~Ticknor, B.~I. Schneider, L.~A. Collins, and
  J.~Burgdörfer.
\newblock \emph{Attosecond {Two}-{Photon} {Interferometry} for {Doubly}
  {Excited} {States} of {Helium}}.
\newblock Physical Review Letters \textbf{107}, 093005 (2011).

\bibitem{PazNagBur2015}
R.~Pazourek, S.~Nagele, and J.~Burgdörfer.
\newblock \emph{Probing time-ordering in two-photon double ionization of helium
  on the attosecond time scale}.
\newblock Journal of Physics B: Atomic, Molecular and Optical Physics
  \textbf{48}, 061002 (2015).

\bibitem{LauBac2003}
S.~Laulan and H.~Bachau.
\newblock \emph{Correlation effects in two-photon single and double ionization
  of helium}.
\newblock Physical Review A \textbf{68}, 013409 (2003).

\bibitem{ZhaPenXu2011}
Z.~Zhang, L.-Y. Peng, M.-H. Xu, A.~F. Starace, T.~Morishita, and Q.~Gong.
\newblock \emph{Two-photon double ionization of helium: {Evolution} of the
  joint angular distribution with photon energy and two-electron energy
  sharing}.
\newblock Physical Review A \textbf{84}, 043409 (2011).

\bibitem{HuColCol2005}
S.~X. Hu, J.~Colgan, and L.~A. Collins.
\newblock \emph{Triple-differential cross-sections for two-photon double
  ionization of {He} near threshold}.
\newblock Journal of Physics B: Atomic, Molecular and Optical Physics
  \textbf{38}, L35 (2005).

\bibitem{KheIva2006}
A.~S. Kheifets and I.~A. Ivanov.
\newblock \emph{Convergent close-coupling calculations of two-photon double
  ionization of helium}.
\newblock Journal of Physics B: Atomic, Molecular and Optical Physics
  \textbf{39}, 1731 (2006).

\bibitem{PalResMcc2009}
A.~Palacios, T.~N. Rescigno, and C.~W. McCurdy.
\newblock \emph{Two-{Electron} {Time}-{Delay} {Interference} in {Atomic}
  {Double} {Ionization} by {Attosecond} {Pulses}}.
\newblock Physical Review Letters \textbf{103}, 253001 (2009).

\bibitem{PalHorRes2010}
A.~Palacios, D.~A. Horner, T.~N. Rescigno, and C.~W. McCurdy.
\newblock \emph{Two-photon double ionization of the helium atom by ultrashort
  pulses}.
\newblock Journal of Physics B: Atomic, Molecular and Optical Physics
  \textbf{43}, 194003 (2010).

\bibitem{PalResMcc2009b}
A.~Palacios, T.~N. Rescigno, and C.~W. McCurdy.
\newblock \emph{Time-dependent treatment of two-photon resonant single and
  double ionization of helium by ultrashort laser pulses}.
\newblock Phys. Rev. A \textbf{79}, 033402 (2009).

\bibitem{NgoManMer2014}
J.~Ngoko~Djiokap, N.~Manakov, A.~Meremianin, S.~Hu, L.~Madsen, and A.~F.
  Starace.
\newblock \emph{Nonlinear {Dichroism} in {Back}-to-{Back} {Double} {Ionization}
  of {He} by an {Intense} {Elliptically} {Polarized} {Few}-{Cycle} {Extreme}
  {Ultraviolet} {Pulse}}.
\newblock Physical Review Letters \textbf{113}, 223002 (2014).

\bibitem{PinLiCol2016}
M.~S. Pindzola, Y.~Li, and J.~Colgan.
\newblock \emph{Multiphoton double ionization of helium using femtosecond laser
  pulses}.
\newblock Journal of Physics B: Atomic, Molecular and Optical Physics
  \textbf{49}, 215603 (2016).

\bibitem{NgoSta2017}
J.~M. Ngoko-Djiokap and A.~F. Starace.
\newblock \emph{Doubly-excited state effects on two-photon double ionization of
  helium by time-delayed, oppositely circularly-polarized attosecond pulses}.
\newblock Journal of Optics \textbf{19}, 124003 (2017).

\bibitem{UllMosDor2003}
J.~Ullrich, R.~Moshammer, A.~Dorn, R.~Dörner, L.~P.~H. Schmidt, and
  H.~Schmidt-Böcking.
\newblock \emph{Recoil-ion and electron momentum spectroscopy:
  reaction-microscopes}.
\newblock Reports on Progress in Physics \textbf{66}, 1463 (2003).

\bibitem{HenEckHar2018_1}
K.~Henrichs, S.~Eckart, A.~Hartung, D.~Trabert, J.~Rist, H.~Sann, M.~Pitzer,
  M.~Richter, H.~Kang, M.~S. Sch\"offler, M.~Kunitski, T.~Jahnke, and
  R.~D\"orner.
\newblock \emph{Experimental evidence for selection rules in multiphoton double
  ionization of helium and neon}.
\newblock Phys. Rev. A \textbf{97}, 031405 (2018).

\bibitem{HenEckHar2018_2}
K.~Henrichs, S.~Eckart, A.~Hartung, D.~Trabert, K.~Fehre, J.~Rist, H.~Sann,
  M.~Pitzer, M.~Richter, H.~Kang, M.~S. Sch\"offler, M.~Kunitski, T.~Jahnke,
  and R.~D\"orner.
\newblock \emph{Multiphoton double ionization of helium at 394 nm: A fully
  differential experiment}.
\newblock Phys. Rev. A \textbf{98}, 043405 (2018).

\bibitem{PinRobLoc2007}
M.~S. Pindzola, F.~Robicheaux, S.~D. Loch, J.~C. Berengut, T.~Topcu, J.~Colgan,
  M.~Foster, D.~C. Griffin, C.~P. Ballance, D.~R. Schultz, T.~Minami, N.~R.
  Badnell, M.~C. Witthoeft, D.~R. Plante, D.~M. Mitnik, J.~A. Ludlow, and
  U.~Kleiman.
\newblock \emph{The time-dependent close-coupling method for atomic and
  molecular collision processes}.
\newblock Journal of Physics B: Atomic, Molecular and Optical Physics
  \textbf{40}, R39 (2007).

\bibitem{ColPin2002}
J.~Colgan and M.~S. Pindzola.
\newblock \emph{Core-{Excited} {Resonance} {Enhancement} in the {Two}-{Photon}
  {Complete} {Fragmentation} of {Helium}}.
\newblock Physical Review Letters \textbf{88}, 173002 (2002).

\bibitem{ResMcc2000}
T.~N. Rescigno and C.~W. McCurdy.
\newblock \emph{Numerical grid methods for quantum-mechanical scattering
  problems}.
\newblock Physical Review A \textbf{62}, 032706 (2000).

\bibitem{MccHorRes2001}
C.~W. McCurdy, D.~A. Horner, and T.~N. Rescigno.
\newblock \emph{Practical calculation of amplitudes for electron-impact
  ionization}.
\newblock Physical Review A \textbf{63}, 022711 (2001).

\bibitem{SchCol2005}
B.~I. Schneider and L.~A. Collins.
\newblock \emph{The discrete variable method for the solution of the
  time-dependent {Schrödinger} equation}.
\newblock Journal of Non-Crystalline Solids \textbf{351}, 1551 (2005).

\bibitem{SchColHu2006}
B.~I. Schneider, L.~A. Collins, and S.~X. Hu.
\newblock \emph{Parallel solver for the time-dependent linear and nonlinear
  Schr\"odinger equation}.
\newblock Physical Review E \textbf{73}, 036708 (2006).

\bibitem{ParLig1986}
T.~J. Park and J.~C. Light.
\newblock \emph{Unitary quantum time evolution by iterative Lanczos reduction}.
\newblock The Journal of Chemical Physics \textbf{85}, 5870 (1986).

\bibitem{SmyEdwPar1998}
E.~S. Smyth, J.~S. Parker, and K.~T. Taylor.
\newblock \emph{Numerical integration of the time-dependent {Schrödinger}
  equation for laser-driven helium}.
\newblock Computer Physics Communications \textbf{114}, 1 (1998).

\bibitem{LefBisCer1991}
C.~Leforestier, R.~H. Bisseling, C.~Cerjan, M.~D. Feit, R.~Friesner,
  A.~Guldberg, A.~Hammerich, G.~Jolicard, W.~Karrlein, H.~D. Meyer, N.~Lipkin,
  O.~Roncero, and R.~Kosloff.
\newblock \emph{A comparison of different propagation schemes for the time
  dependent {Schrödinger} equation}.
\newblock Journal of Computational Physics \textbf{94}, 59 (1991).

\bibitem{FouAntBac2008}
E.~Foumouo, P.~Antoine, H.~Bachau, and B.~Piraux.
\newblock \emph{Attosecond timescale analysis of the dynamics of two-photon
  double ionization of helium}.
\newblock New Journal of Physics \textbf{10}, 025017 (2008).

\bibitem{MorWatLin2007}
T.~Morishita, S.~Watanabe, and C.~D. Lin.
\newblock \emph{Attosecond Light Pulses for Probing Two-Electron Dynamics of
  Helium in the Time Domain}.
\newblock Phys. Rev. Lett. \textbf{98}, 083003 (2007).

\bibitem{JiaWeiTia2014}
W.-C. Jiang, W.-H. Xiong, T.-S. Zhu, L.-Y.~P. Peng, and Q.~Gong.
\newblock \emph{Double ionization of He by time-delayed attosecond pulses}.
\newblock Journal of Physics B: Atomic, Molecular and Optical Physics
  \textbf{47}, 091001 (2014).

\bibitem{HorMorRes2007}
D.~A. Horner, F.~Morales, T.~N. Rescigno, F.~Mart\'{\i}n, and C.~W. McCurdy.
\newblock \emph{Two-photon double ionization of helium above and below the
  threshold for sequential ionization}.
\newblock Phys. Rev. A \textbf{76}, 030701 (2007).

\end{thebibliography}
\end{document}